\documentstyle[aps,prb]{revtex}
\begin{document}
\draft
\title{Photon-Assisted Quasiparticle 
Transport and Andreev Transport through an 
       Interacting Quantum Dot}
\author{Sam Young Cho$^1$, Kicheon Kang$^{1,2}$,
and Chang-Mo Ryu$^1$}
\address{$^1$Department of Physics, 
Pohang University of Science and Technology,
	 Pohang 790-784, Korea}
\address{$^2$Max-Planck-Institut f\"ur Physik Komplexer Systeme
  N\"othnitzer Strasse 38, D-01187 Dresden, Germany}
\date{\today}
\maketitle

\begin{abstract}
Resonant tunneling through a quantum dot coupled to
superconducting reservoirs in the presence of
time-dependent external voltage has been studied.
A general formula of the current is derived based 
on the nonequilibrium Green's function technique. 
Using this formula photon-assisted quasiparticle transport 
has been investigated for the 
quantum dot connected to superconductors.
In addition, resonant Andreev transport through
a strongly correlated quantum dot connected to 
a normal metallic lead and a superconducting lead is  studied. 
\end{abstract}


%
\section{Introduction}
Recent advance of nano-technology has stimulated much
interest in the study of quantum transport 
in mesoscopic structures.
Among the various mesoscopic structures, 
devices based on quantum dot ({\em QD}) have 
drawn  particular attention\cite{Kouwen97}.
For the quantum dot structures, electron transport is shown to be
affected by the confined electrons in the dot.
Very recently, novel {\em tunable} Kondo effect has been 
experimentally found in 
a single electron transistor\cite{Goldhaber,cronenwett98},
where the level position of the quantum dot and the tunneling
rate were controlled by the external gate voltages,
via Coulomb blockade.

The influence of time-varying fields on the transport
through a quantum dot structure is one of potentially 
interesting areas.
Particular applications related to this physics 
are photo-electron devices, such as single electron pumps,
turnstiles, and photon detectors 
(see Ref.\onlinecite{Kouwen97} for a review).
A time-dependent potential with frequency $\Omega$
superimposed to dc bias potential
can induce additional tunneling processes when electrons exchange
energy by absorbing or emitting photons of energy $\Omega$.
This kind of tunneling is known as 
the photon-assisted tunneling (PAT)\cite{Tien}.
Experimental study on the PAT in the quantum dot devices, 
based on a configuration of
a quantum dot  coupled to normal metal leads ({\em N}),
has been reported \cite{Kouwen,Blick,Fuji,Oost}.
 
In this paper we wish to study tunneling properties 
in an interacting quantum dot coupled to superconductors 
in the presence of an external ac voltage.
For this purpose we first derive 
a general current formula for this system.
Using this formula we investigate two examples
of time-dependent transport,
PAT of quasiparticles and that
of the Kondo-Andreev resonance.
When the quantum dot is coupled to a superconductor ({\em S}),
the resonant tunneling current is strongly  affected by the
singular BCS density of states\cite{Tinkhams}.
We find that the singularity of BCS density of states plays 
an important role in the photon assisted quasiparticle transport.
If the coupling between the leads and the {\em QD} is sufficiently
weak, the subgap transport is suppressed in the case of 
small dc bias, but the photon assisted tunneling combined
with the  asymmetric BCS density of states could 
allow finite electron transport.
We also investigate the many-body resonance  
by an ac applied voltage in the {\em N-QD-S} system 
as an application of the derived current formula.
We find new side peaks
related to the photon absorption 
and/or emission of the strongly correlated electrons,
in addition to the zero-bias peak due to the Kondo resonance
with the Andreev reflection\cite{Andreev}.
 
The paper is organized as follows.
In Sec. II we derive a  time-dependent current formula  
for the quantum dot
connected to superconducting  leads.  
This is an extension of the
current formula for the normal metallic leads, 
derived by Jauho, Wingreen, and Meir \cite{Wing1}.
On this  base, the photon-assisted
quasiparticle pumping current is studied in Sec. III.
The Andreev current for photon-assisted tunneling
is studied in Sec. IV. 
Main results of this paper   are summarized in Sec. V.

\section{Current formula}
We start with a general model Hamiltonian
\begin{equation}
H \!=\! \sum\nolimits_{\alpha} \!\! H_{\alpha} 
\{ {c}_{k\sigma}^{\dagger};{c}_{k\sigma} \}
+ \sum\nolimits_{\alpha} \!\! H^{\alpha}_T 
+ H_D \{ {d}_\sigma^{\dagger};{d_\sigma} \} ,
\label{hamiltonian}\end{equation}
where $\alpha(\in \lbrace L,R \rbrace, 
\mbox{denoting the left hand and 
right hand sides})$ is the index of the leads.         
$H_\alpha$, $H_D$, and 
$ H^{\alpha}_T= \sum_{k\sigma\in {\alpha}}
[ V^{\alpha}_{k\sigma}{{c}}_{k\sigma}^{\dagger} 
{{d}}_\sigma + {\mathrm h.c.}]$
represent the Hamiltonians of superconducting (or normal) 
lead, an interaction region, and tunneling between the lead 
and the interaction region, respectively.  
The superconducting lead $\alpha$ is characterized 
by the chemical potential $\mu_\alpha$ 
and the gap energy $\Delta_\alpha$.
${d}^\dagger_\sigma $ ($ {d}_\sigma $) is the
electron creation (annihilation) operator with spin $\sigma$
in the quantum dot.
When there is a time-dependent voltage difference
 $eV^\alpha(t)=\mu_\alpha(t)-\mu_D(t)$ 
between the lead $\alpha$ and the quantum dot,
it is convenient to perform gauge transformation \cite{Rogo}.
The electron creation and annihilation operator 
for the lead takes the form
${\hat c}_{k\sigma}^{\dagger} ={\mathrm e}^{i\phi_\alpha/2}
{c}_{k\sigma}^\dagger$ and
${\hat c}_{k\sigma} ={\mathrm e}^{-i\phi_\alpha/2}
{c}_{k\sigma}$,
where $d\phi_\alpha/dt= 2eV^\alpha/\hbar$.

The current flowing into the quantum dot 
can be defined as the rate of change in the number of electrons
in a lead $\alpha$.
The commutator of the number operator 
$N_\alpha = \sum_{k\sigma \in \alpha}
 c^\dagger_{k\sigma} c_{k\sigma} $ 
with the Hamiltonian (\ref{hamiltonian}) gives rise to the current,
\begin{equation}
J_{\alpha}(t)
 = -\frac{ie}{\hbar}
    {\mathrm Re} \Bigl\lbrace \sum_{k\sigma \in \alpha}
          V^*_{k\sigma} \langle d^\dagger_\sigma {\hat c}_{k\sigma} 
          \rangle \Bigr\rbrace .
\end{equation}
In the case of superconducting lead,  several kinds of 
tunneling processes, i.e, quasiparticles, Cooper-pairs, etc.,
can be included in the general current expression. 
In order to treat the system in a convenient way,
we have adopted the particle-hole Nambu representations 
in the lead and the quantum dot such as 
${\mathbf \Psi}^{\dagger}_{k\sigma}(t) = 
( {\hat c}^{\dagger}_{k\sigma}(t)\; {\hat c}_{-k-\sigma}(t) )$ 
and
${\mathbf \Psi}^{\dagger}_\sigma (t)= 
( {d}^{\dagger}_{\sigma} (t) \; {d}_{-\sigma} (t) )$, respectively.
In the particle-hole space, new Green's function can be defined 
in terms of Keldysh Green's function,
\begin{mathletters}
\begin{eqnarray}
{\mathbf G}^{<}_{k\sigma\!,d}(t,t') \! 
 &=&i \langle {\mathbf \Psi}^{\dagger}_\sigma (t')
             \otimes {\mathbf \Psi}_{k\sigma} (t) \rangle \\
 &=&i\!\! \left( \! \! \begin{array}{cc}
    \! \langle \ d^\dagger_\sigma(t') \; {\hat c}_{k\sigma}(t) \ \rangle \! &
    \! \langle \ d_{-\sigma}(t') \; {\hat c}_{k\sigma}(t)\  \rangle \! \\
    \! \langle d^\dagger_\sigma(t'){\hat c}^\dagger_{-k,-\sigma}(t) \rangle
\! &
    \! \langle d_{-\sigma}(t') {\hat c}^\dagger_{-k,-\sigma}(t) \rangle
	      \end{array} \!\!  \right).
\end{eqnarray}
\end{mathletters}
The time-dependent current flowing out of the lead $\alpha$ into 
the quantum dot can be written as
\begin{equation}
J_{\alpha}(t) = -{2e\over\hbar} {\mathrm{ Re}}
\Bigl\lbrace \sum_{k\sigma\in {\alpha}}
{\mathbf V}^{\alpha *}_{k\sigma}
{\mathbf G}_{k\sigma,d}^<(t,t)\Bigr\rbrace_{11} ,
\end{equation}
where $\lbrace \cdots \rbrace_{11}$ stands for the $(1, 1)$ 
component of the current matrix and the hopping matrix is 
$ {\mathbf V}^\alpha_{k\sigma}= 
  \left( \begin{array}{cc} V^\alpha_{k\sigma} & 0 \\ 
                           0  &- V^{\alpha *}_{-k-\sigma}
          \end{array} \right)$. 

By means of Keldysh technique\cite{Keldysh}
for the noninteraction Hamiltonian of the leads, 
we have obtained  Dyson's equation
\begin{mathletters}
\begin{eqnarray}
{\mathbf G}^{<}_{k\sigma,d}(t,t')
&=&i\langle T_C [ S {\mathbf \Psi}^{\dagger}_{\sigma}(t') 
\otimes {\mathbf \Psi}_{k\sigma}(t)] \; \rangle \\
&=&\int dt_1 \Bigl( {\mathbf g}^r_{k\sigma}(t,t_1) 
{\mathbf V}^\alpha_{k\sigma} {\mathbf G}^<_\sigma(t_1,t') +
{\mathbf g}^<_{k\sigma}(t,t_1) {\mathbf V}^\alpha_{k\sigma} 
{\mathbf G}^a_\sigma(t_1,t') \Bigr) \; ,
\end{eqnarray}
\end{mathletters}
where
$S = T_C \{ \exp[-i\int_C d\tau H_T(\tau)]\}$ is 
the contour-ordered $S$-matrix and $T_C$
is the contour-ordering operator.
${\mathbf G}^{<}_\sigma(t,t') =i \langle 
{\mathbf \Psi}^{\dagger}_\sigma (t')
\otimes {\mathbf \Psi}_{\sigma} (t)\rangle$
and
${\mathbf G}^{a}_\sigma(t,t') = i \theta(t'-t)
\langle \{ {\mathbf \Psi}_\sigma (t), 
{\mathbf \Psi}^{\dagger}_{\sigma} (t') \}\rangle$ 
are the lesser and  advanced Green's matrices 
in the interaction region. 
The full Green's functions ${\mathbf G}_\sigma$ 
in the  interaction region need to be solved.
The Green's function  ${\mathbf g}_{k\sigma}$
represents the unperturbed Green's function in the lead, which 
includes the time-dependent phase due to the chemical potential
difference and the electron level variation.

In order to obtain the time-dependent Green's functions
in the superconducting leads, we take out 
the time-dependent phase associated with a voltage difference in
the electron operators as we mentioned above.
Assuming $\Delta_\alpha$ to be independent on the external
fields in the superconducting leads, we can separately consider 
the time variations of the energy levels in the leads 
and in the  interaction region  
(for example, $\varepsilon_k(t)=\varepsilon_k+ V^\alpha_{ac}
\cos\Omega t$ ).
The total phase $\phi_\alpha(t)$  includes the phase due to
the time variation of the energy levels. 
The Green's functions of a superconducting lead 
have a form\cite{Artem} 
such as ${\mathbf g}_{k\sigma}(t,t')={\mathbf R}_z(\phi_\alpha(t))
{\mathbf g}_{k\sigma}(t-t'){\mathbf R}^*_z(\phi_\alpha(t'))$,
where
${\mathbf R}_z(\phi_{\alpha})={\mathrm exp}[-i (\phi_{\alpha}/2) 
\bbox{\sigma}\!_z]$ is a rotation operator in the two-dimensional
complex space.
The electron operators $c$ transform to the quasi-particle operators 
$\gamma$ by means of the Bogoliubov-Valatin transformation\cite{Bogol},
\begin{equation}
\left( \begin{array}{c} c_{k\sigma} \\ c^\dagger_{-k,-\sigma}
       \end{array} \right)
=\left( \begin{array}{cc} u_k &  v_k \\
                         -v_k &  u_k 
        \end{array} \right)
\left( \begin{array}{c} \gamma_{k\sigma} \\ \gamma^\dagger_{-k,-\sigma}
       \end{array} \right)
\end{equation}
with  the BCS coherence factors $u_k$ and $v_k$.
The time-dependent Green's functions in the leads
can be written as 
\begin{mathletters}
\begin{eqnarray}
{\mathbf g}^{r}_{k\sigma}(t,t')&=&-i \theta(t-t')
{\mathbf R}_z(\phi_{\alpha}(t)) 
\left[ {\mathrm e}^{-i E_k(t-t')}
{\mathbf U}^\alpha_{\mathbf k}
+{\mathrm e}^{i E_k (t -t')}
{\mathbf V}^\alpha_{\mathbf k} \right]
{\mathbf R}^*_z(\phi_{\alpha}(t') ) ,
\\
{\mathbf g}^<_{k\sigma}(t,t')&=& i \,
{\mathbf R}_z(\phi_{\alpha}(t)) 
\left[ f_{\alpha}(E_k) {\mathrm e}^{-i E_k(t-t')}
{\mathbf U}^\alpha_{\mathbf k}
+[1-f_{\alpha} (E_k)] {\mathrm e}^{i E_k (t-t')}
{\mathbf V}^\alpha_{\mathbf k} \right]
{\mathbf R}^*_z(\phi_{\alpha}(t')) ,
\end{eqnarray}
\end{mathletters}
where 
$f_{\alpha}(E_k)$ is the Fermi distribution function 
in the lead $\alpha$. 
One can define the matrix coherence factors 
$\mathbf U^\alpha_k$ and $\mathbf V^\alpha_k$  given by
\begin{equation}
\begin{array}{cc}
{\mathbf U^\alpha_k}=\left( \begin{array}{cc}
	u_k^2 & -u_k v_k \\ 
	-u_k v_k & v_k^2 \end{array} \right) \mbox{~~ and~~ }&
{\mathbf V^\alpha_k}=\left( \begin{array}{cc}
	 v_k^2 &  u_k v_k \\ 
	 u_k v_k & u_k^2 \end{array} \right).
\end{array}
\end{equation}
The elements of the matrix coherence factors are given by
$u^2_k=1/2( 1 + \varepsilon_k /E_k )$ and 
$v^2_k=1/2( 1 - \varepsilon_k / E_k)$.
Here, $E_k=\sqrt{\varepsilon_k^2+\mid\Delta_\alpha\mid^2}$.
For normal metallic leads $(\Delta_\alpha=0)$, 
$\varepsilon_k$ becomes the  electron energy,
and the matrix coherence factors are reduced to 
${\mathbf U^\alpha_k}=\theta(\varepsilon_k){\mathbf P\!_+}
+\theta(-\varepsilon_k) {\mathbf P\!_-}$ and
${\mathbf V^\alpha_k}=\theta(-\varepsilon_k){\mathbf P\!_+}
+\theta(\varepsilon_k) {\mathbf P\!_-}$,
where ${\mathbf P}\!_{\pm} = 1/2({\mathbf 1} \pm 
\bbox{\sigma}\!_z )$ are the polarization matrices
and $ \bbox{\sigma}\!_z$($\mathbf 1$)
represent the Pauli (unit) matrices.
%
The time-dependent coupling matrix
can be  written in a compact form as 
${\mathbf \Gamma}^\alpha_S(E_k,t,t')=2\pi 
N_S^\alpha(E_k) 
{\mathbf V}^{\alpha *}_{k\sigma} {\mathbf R}_z(t) 
{\mathbf W^\alpha_ k} {\mathbf R}^*_z(t')
{\mathbf V}^{\alpha}_{k\sigma}$,
where 
${\mathbf W^\alpha_k}(E_k)={\mathbf U^\alpha_k}
(\varepsilon_k,E_k)+{\mathbf V^\alpha_k}(-\varepsilon_k,-E_k)$
and 
$N^\alpha_S(E_k)$ is the density of states  
of the superconducting lead $\alpha$.
As a consequence, the time-dependent current for 
the systems coupled to two  superconducting leads can be written as 
\begin{equation}
J_{\alpha}(t)\!= \! {2e\over\hbar}\!
\int_{-\infty}^t \!\!\! dt_1 \!\! \int \! \frac{dE_k}{2\pi} 
{\mathrm Im} \Bigl\lbrace \sum_{\sigma} 
{\mathrm e}^{-i E_k (t-t_1\!)} 
{\mathbf \Gamma}^\alpha_S(E_k,t,t_1)[-{\mathbf G}^<_\sigma(t_1,t)
+f_{\alpha}(E_k){\mathbf G}^a_\sigma(t_1,t)] \Bigr\rbrace_{11},
\label{formula}\end{equation}
with the time-dependent coupling matrix 
\begin{equation}
{\mathbf \Gamma}^\alpha_S(E,t,t')=\frac{ {\mit \Gamma}^\alpha }
{\sqrt{E^2 -\Delta_\alpha^2}}
\left( \begin{array}{cc}
        {\mathrm e}^{-i\phi(t)/2} & 0 \\
        0 & {\mathrm e}^{i\phi(t)/2}
	\end{array} \right)
\left( \begin{array}{cc}
       \mid E \mid & {\mathrm sign}(E) \Delta_\alpha  \\
       {\mathrm sign}(E) \Delta_\alpha & \mid E \mid
       \end{array}
\right)
\left( \begin{array}{cc}
        {\mathrm e}^{i\phi(t')/2} & 0 \\
        0 & {\mathrm e}^{-i\phi(t')/2}
	\end{array} \right) .
\end{equation}
The coupling constant, which is assumed to be independent of energy
and spin, is defined by ${\mit \Gamma}^{\alpha}
=2\pi \mid V^\alpha_{k\sigma} \mid^2 N^\alpha(0)$
with $N^\alpha(0)$ being the normal density of states at the Fermi
level. 

In the limit $\Delta_\alpha=0$,
the current expression given in Eq. (\ref{formula}) reduces to
the formula obtained by  Jauho, Wingreen and Meir\cite{Wing1} 
for normal metallic leads.
In principle, the current formula of Eq.(\ref{formula})
can describe various transport processes, 
associated with the superconducting leads.
The time dependent transport physics
for a quantum dot connected to the superconducting leads
with the  applied time-dependent voltage
can be  studied based on  Eq. (\ref{formula}) once 
the interaction in the quantum dot is determined. 

\section{Photon-assisted quasiparticle current}
On the basis of the formula derived in the previous section,
we now study the resonant quasiparticle 
tunneling process
in the presence of an oscillating external ac voltage.
We assume that the quantum dot  is weakly coupled to the 
superconducting leads.
The Coulomb charging energy $( U \sim e^2/C )$, where $C$
is the total capacitance of the quantum dot, and the level
spacing of the discrete electronic states are much greater
than the energy gap.
Therefore, in the weak coupling limit 
$\Gamma \ll \Delta_\alpha \ll U$,
the subgap transport due to Andreev reflection is negligible 
due to large charging energy 
in the quantum dot\cite{Tinkhams,Yeyati,Kckang}.
%
Recently, Oosterkamp, Kouwenhoven, Koolen, Vaart, and Harmans\cite{Oost} 
observed photon induced pumping of the dc current via the 0D state 
in the normal single electron transistor. 
They also observed  photon sideband resonances. 
Compared to the case of normal metallic leads, 
here in the case of the superconducting leads much enhanced 
currents are shown in the photon-induced pumping
and the direction of the current changes,
due to the  singular BCS density of states.

We have calculated the time-averaged current,
in the presence of sinusoidal external voltage difference between
the lead $\alpha$ and the quantum dot with the frequency $\Omega$ 
and the oscillation amplitude $V_{ac}^{\alpha}$.
By neglecting the off-diagonal terms of the general current 
expression which are related to the Andreev transport,
and averaging with respect to time, 
the dc current  can be written as 
\begin{equation}
J^Q_{\alpha}\!=\!\!{ie\over \hbar} \!\sum_\sigma \!\! \int\! 
\frac{d\varepsilon}{2\pi} \Big\{{\mit \Gamma}^\alpha_S (\varepsilon) 
{\cal G}^<_\sigma(\varepsilon)\! +\! K^\alpha_S(\varepsilon) \!
\left( {\cal G}^r_\sigma(\varepsilon)\!-\!{\cal G}^a_\sigma
(\varepsilon) \right)\!\! \Big\},
\end{equation}
where ${\cal G}$ denotes the $(1,1)$ component of the
Green's function matrix ${\mathbf G}$.
${\mit \Gamma}_S^\alpha(\varepsilon)=\int d\varepsilon' 
\, {\mit \Gamma}^\alpha \, \varrho_S^\alpha(\varepsilon') \,
Q^{-}_\alpha(\varepsilon-\varepsilon')$ and 
$K_S^\alpha(\varepsilon)=\int d\varepsilon' 
\, {\mit \Gamma}^\alpha \, \varrho_S^\alpha(\varepsilon') \,
f_\alpha(\varepsilon') \, Q^{-}_\alpha(\varepsilon-\varepsilon')$.
The dimensionless BCS factor is
$\varrho^S_\alpha(\varepsilon)=
\mid \varepsilon \mid /\sqrt{\varepsilon^2 - \mid\Delta_\alpha\mid^2}$.
The effects of photon absorption and emission processes
are included in the probability function
$Q^{\pm}_\alpha(\varepsilon)=\sum_n J^2_n(\lambda_\alpha) 
\delta(\varepsilon \pm \mu_\alpha \pm n\Omega)$ where 
$J_n(\lambda_\alpha)$ is the $n$-th order Bessel function 
with the argument of the normalized oscillation amplitude
$\lambda_\alpha={\mit eV_{ac}^\alpha}/\hbar\Omega$.

To describe the interacting quantum dot we consider
the Anderson impurity model 
$ H_D=\sum_\sigma \varepsilon_\sigma {d}^\dagger_\sigma {d}_\sigma 
+ U n_\uparrow n_\downarrow$ 
for $\varepsilon_\sigma = \varepsilon_{-\sigma}$.
In general, the Green's functions of the quantum dot can be
found from Dyson's equation  
${\cal G}^{r,a}={\cal G}^{r,a}_0 [ 1 + \Sigma^{r,a} {\cal G}^{r,a} ]$
and ${\cal G}^< = [ 1 + {\cal G}^r \Sigma^r ]
{\cal G}^<_0 [ 1 + \Sigma^a {\cal G}^a ] + 
{\cal G}^r \Sigma^< {\cal G}^a$. 
By using the equation of motion technique and 
the mean field approximation\cite{Zhao},
for the photon-assisted tunneling, we obtain
the approximate Green's function  
$
{\cal G}^<_\sigma(\varepsilon) \simeq 
-( K_S(\varepsilon)/{\mit \Gamma}_S (\varepsilon) )
\big( {\cal G}^r_\sigma(\varepsilon) 
- {\cal G}^a_\sigma(\varepsilon) \big),
$
where ${\mit \Gamma}_S(\varepsilon)=\sum_\alpha 
{\mit \Gamma}_S^\alpha(\varepsilon)$ and
$K_S(\varepsilon)=\sum_\alpha K_S^\alpha(\varepsilon)$.
This approximation guarantees automatically the current conservation.
The time averaged current through the system becomes
\begin{equation}
J^Q=\frac{e}{\hbar} \sum_\sigma \int d\varepsilon \,
{\widetilde{\mit \Gamma}}(\varepsilon) \, A^Q_\sigma(\varepsilon),
\label{quasi}\end{equation}
where ${\widetilde{\mit \Gamma}}(\varepsilon) \equiv 
\left(K^L_S (\varepsilon){\mit \Gamma}^R_S(\varepsilon)
-K^R_S(\varepsilon) {\mit \Gamma}^L_S(\varepsilon)\right)/
{\mit \Gamma}_S(\varepsilon)$ and
$A^Q_\sigma(\varepsilon)=-{\mathrm Im}
\left[{\cal G}^r_\sigma(\varepsilon)\right]/\pi$.
In the absence of the ac voltage, 
Eq.(\ref{quasi}) reduces to the current
formula derived in Ref.\onlinecite{Kckang}
for the quasiparticle resonant tunneling of a quantum dot
connected to superconductors.
After neglecting the higher-order correlation functions\cite{Haug},
the approximate Green's function ${\cal G}^r_\sigma(\varepsilon)$
can be written as
\begin{equation}
{\cal G}^r_\sigma(\varepsilon)=
\frac{1-\langle n_{-\sigma} \rangle}
{\varepsilon-\varepsilon_\sigma-\Sigma^r_0(\varepsilon)}
+ \frac{\langle n_{-\sigma} \rangle}
{\varepsilon-\varepsilon_\sigma-U-\Sigma^r_0(\varepsilon)},
\end{equation}
where the self-energy is 
$\Sigma^r_0(\varepsilon)=-i{\mit \Gamma}_S(\varepsilon)/2$.
Here we do not take into account the Kondo-like correlations 
because it is not important in the case of
both leads being superconductors.
The occupation number
in the nonequilibrium state can be obtained from the relation 
$\langle n_{-\sigma} \rangle = (1/2\pi i) \int d\varepsilon 
\; {\cal G}^<_\sigma(\varepsilon)$.
Then, the current and the occupation number
of the quantum dot are solved self-consistently.

\begin{figure}
\vspace*{15.5cm}
\includegraphics{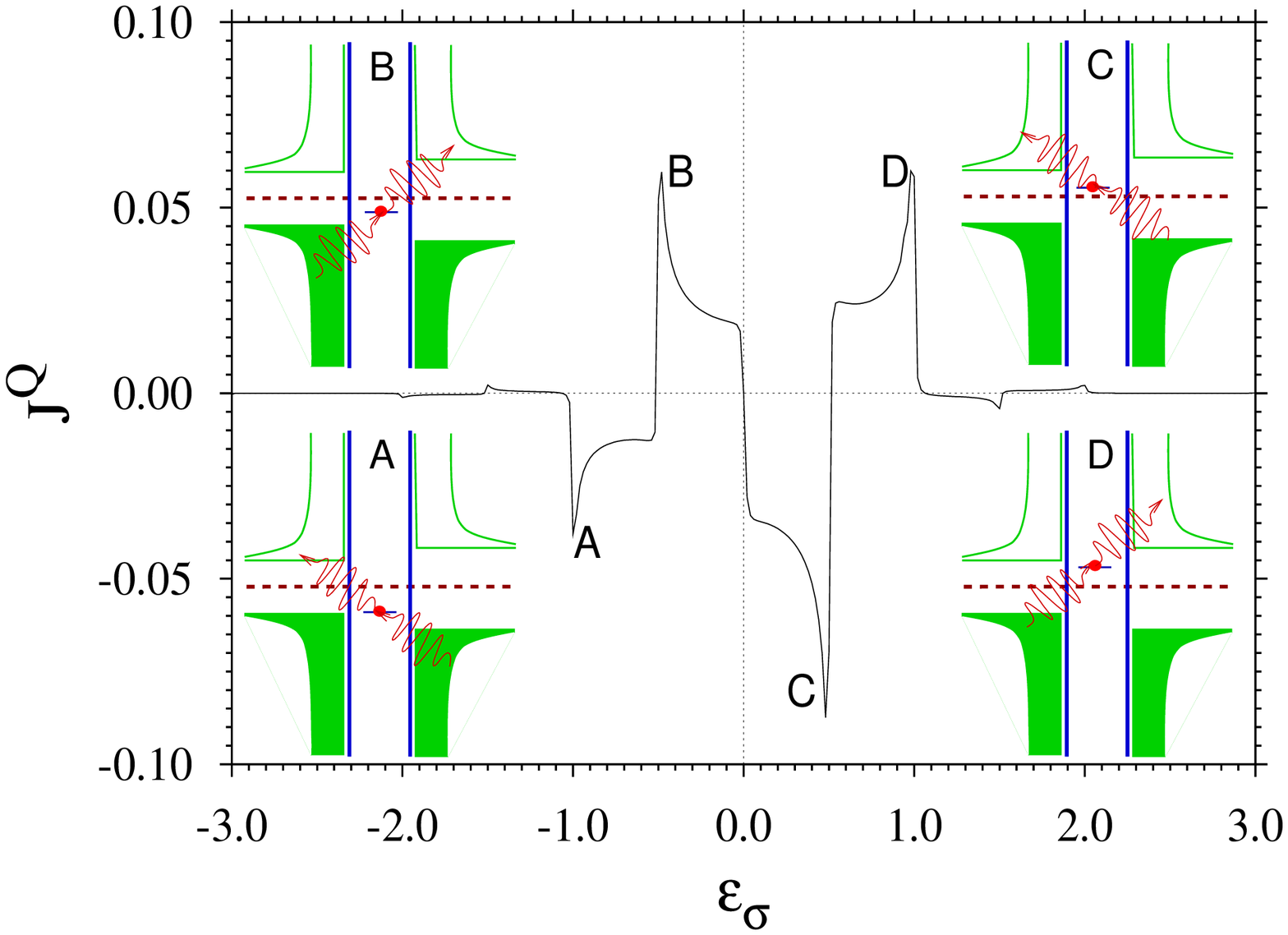}
\caption{
Zero-bias quasiparticle current 
of the {\em S-QD-S} in units of $e{\mit \Gamma}/\hbar$ 
for asymmetric gap energy; $\Delta_L=1.0$, $\Delta_R=1.5$. 
The other parameters taken here are
 ${\mit \Gamma}={\mit \Gamma}^L={\mit \Gamma}^R=0.01$,
$k_BT=0.1$, $U=20$, $\lambda_L=\lambda_R=1.0$, $\Omega=1.0$, in
units of $\Delta_L$.
Diagrams of the inset displays the dominant two photon
processes for several cases
denoted by $\mathsf A, B, C$ and $\mathsf D$.
 }
\label{fig1}\end{figure}

When the quantum dot is coupled to the superconductors,
the pumping current due to quasiparticles can flow even 
when  the ac-fields are applied to both sides of the leads.  
In the case of the normal lead, on the other hand,
the electron pumping current can flow only when
the ac-field is applied between the dot and one lead \cite{Oost,Bruder}.
In the normal lead case, when the ac-fields are applied to both sides of
the leads,
the current can not flow.
Figure \ref{fig1} displays the zero-bias current  
for unequal superconducting gap energies in the two leads.
In this case, since the external frequency is smaller 
than the gap, the single photon processes are suppressed, and
two photon processes  become the  dominant one 
because of the BCS gap.
The sharp peak of the  
current induced by the oscillating
field reflects the singularity of the BCS density of states.
The sharp peaks can give rise to a good spectroscopic resolution.
This suggests that a  {\em S-QD-S} system  can be utilized
as a potentially good photo-electron device.
Electron transmission  processes are
depicted in the inset of the Fig.\ref{fig1}.
Note that the direction and the magnitude of the current change 
as the level position in the quantum dot varies. 

\begin{figure}
\vspace*{15.5cm}
\includegraphics{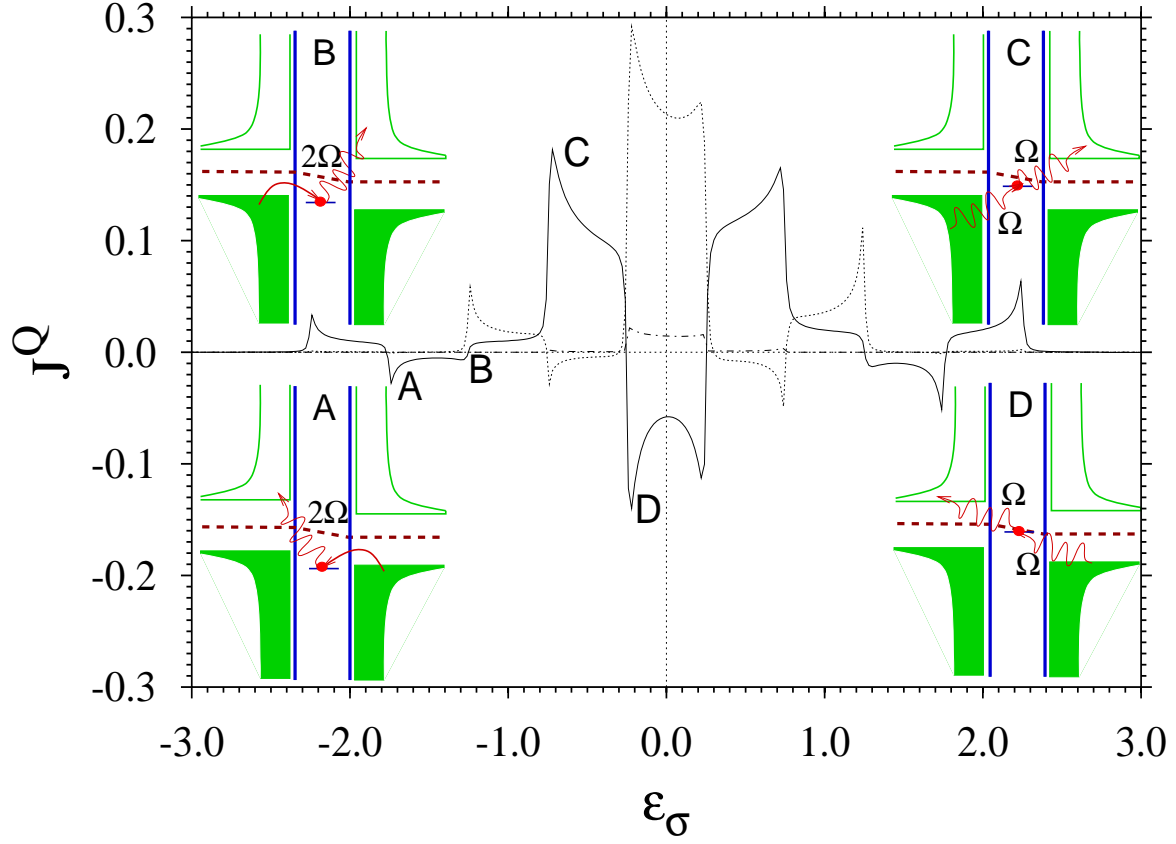}
\caption{
Photon induced pumping of the quasiparticle current 
for the {\em S-QD-S} in units of $e{\mit \Gamma}/\hbar$
for a small dc bias with two identical superconductors.
$\mu_L=-\mu_R=eV/2$, $eV=0.5$, 
${\mit \Gamma}={\mit \Gamma}^L={\mit \Gamma}^R=0.01$,
$k_BT=0.1$, $U=20$, $\lambda_L=\lambda_R=1.0$, 
$\Omega=1.5$(solid line), $\Omega=1.0$(dotted line), 
and $\Omega=0.5$(dash-dotted line) in units of 
$\Delta=\Delta_L=\Delta_R$.
Dominant processes with $\Omega=1.5$ are depicted 
in the diagrams of the
inset for several values of the energy level in the {\em QD}.}
\label{fig2}\end{figure}

Figure \ref{fig2} shows photon pumping of 
the current for two identical superconductors,
when a small dc voltage is applied to the system.
One can clearly see the negative current in some regions 
of parameters ($\mathsf{A,D}$), 
which means the negative differential conductance
arises due to the photon pumping of the 
current.
The dominant tunneling processes for $\Omega=1.5\Delta$ are depicted
in the inset for several values of $\varepsilon_\sigma$.
In  regions $\mathsf{A}$ and $\mathsf{B}$
 the two-photon process becomes  dominant, whereas a single photon 
photon processes dominates, giving rise to a greater current,
if the dot level lies in between the BCS gaps ($\mathsf{C,D}$).
It is noteworthy that the direction and the
amplitude of the current
change depending on the external frequency. As one can see 
in the Fig.\ref{fig2}, the direction of the current 
differs for different frequencies of  $\Omega=0.5\Delta$, 
$1.0 \Delta$ and $1.5\Delta$.

\section{Photon-assisted Andreev Current}
In the present section we will investigate
the Andreev resonant tunneling through a strongly correlated
quantum dot coupled to a normal metal and to a
superconductor \cite{Fazio,kang98-2}.
We consider that an ac voltage is applied between 
the normal metallic lead and the quantum dot only. 
Generalizing  the {\it ansatz} of Ng\cite{Ng}
to the case of  a superconducting lead\cite{Fazio},
(which leads to current conservation,)
the time averaged current in the presence of sinusoidal 
external ac voltage on the normal lead can be obtained
from  Eq.(\ref{formula})
considering the Andreev reflections, 
\begin{equation}
J^{A}_\alpha=\frac{ie}{\hbar}\!\int \!\frac{d\varepsilon}{2\pi}
{\mathrm Tr}\Big\{ {\bbox \sigma}\!_z 
{\mathbf G}^r_\sigma(\varepsilon)
[{\bbox \Sigma}^r(\varepsilon),
{\mathbf K}^\alpha_N(\varepsilon)]
{\mathbf G}^a_\sigma(\varepsilon) \Big\},
\label{eqAnd}\end{equation}
where 
$ {\mathbf K}^\alpha_N(\varepsilon)=
\left( \begin{array}{cc}
       {K^-_N}(\varepsilon) & 0 \\
       0 & {K^+_N}(\varepsilon)
       \end{array}
\right) $
with ${K^{\pm}_N}(\varepsilon) =\int d\varepsilon' 
\, {\mit \Gamma}^\alpha \, 
f_\alpha(\varepsilon') \, 
Q^{\pm}_\alpha(\varepsilon-\varepsilon')$.
In obtaining the Andreev current we have assumed
that the ac voltage difference is applied only on the normal side
and that the noninteracting self-energies are  in 
the limit $\Omega, k_BT, eV \ll \Delta, U$.
In the absence of the ac voltage, our equation of (\ref{eqAnd}) 
can be shown to be reduced to the current formula derived 
by Fazio and Raimondi\cite{Fazio}. 
The Green's functions can be solved by the equation of motion
technique, taking into account Kondo-like 
correlations\cite{Meir93}.
In the infinite $U$ limit the Green's function matrix 
has the form
\begin{equation}
\left( \bbox{\cal E} -\bbox{\cal E}_\sigma 
-\bbox{\Sigma}^r (\varepsilon) \right) 
{\mathbf G}^r_\sigma (\varepsilon)=\bbox{1} -\langle \bbox{n} \rangle ,
\end{equation}
where 
$\bbox{\cal E}_\sigma=
\left( \begin{array}{cc}
       \varepsilon_\sigma & 0 \\
       0 & -\varepsilon_{-\sigma}
	\end{array} \right) $ 
and
the self-energy can be written as ${\mathbf \Sigma}(\varepsilon)=
{\mathbf \Sigma}^r_N(\varepsilon) + {\mathbf \Sigma}^r_S(\varepsilon)$,
with ${\mathbf\Sigma}^r_N$ and 
${\mathbf\Sigma}^r_S$ denoting the self-energy
contribution from the normal metal and the superconductor, respectively.
The effects of the PAT and Kondo correlations are included
in ${\mathbf\Sigma}^r_{N,11}(\varepsilon)$ as \cite{Ng}
\begin{equation}
{\mathbf\Sigma}^r_{N,11} \!(\varepsilon) \!
= \!\!\! \!
\sum^\infty_{n=-\infty} \!\!\! J^2_n(\lambda_N) \!
\int \!\! \frac{d\varepsilon_k}{2\pi} \!
\frac{ {\mit \Gamma}^N [ 1 + f_N(\varepsilon_k) ] }
{ \varepsilon \!- \!\varepsilon_k \!- \!eV \!- \!n\Omega \!+ \!i0^+}.
\end{equation}
We can confirm the time reversal symmetry that 
${\mathbf \Sigma}^r_{N,22}(\varepsilon)
=-{\mathbf \Sigma}^{r *}_{N,11}(-\varepsilon)$.
In the large $\Delta$ limit ($ \Omega,\varepsilon 
\ll \Delta $), ${\mathbf \Sigma}^r_S$ 
can be written as \cite{Fazio}
\begin{equation}
{\mathbf \Sigma}^r_S =
\left( \begin{array}{cc}
  {\mit \Gamma}^S \ln[W/\Delta]/2\pi & -{\mit \Gamma}^S/2 \\
  -{\mit \Gamma}^S/2 & -{\mit \Gamma}^S \ln[W/\Delta]/2\pi 
	\end{array}
\right).
\end{equation}
The off-diagonal 
terms are directly related to the Andreev transport,
whereas the effect of the diagonal term renormalizes
the energy level of the quantum dot
as $\widetilde \varepsilon_{\sigma,-\sigma}=
\varepsilon_{\sigma,-\sigma}+{\mit \Gamma}^S\ln[W/\Delta]/2\pi$
in the large $\Delta$ limit.

\begin{figure}
\vspace*{15.5cm}
\includegraphics{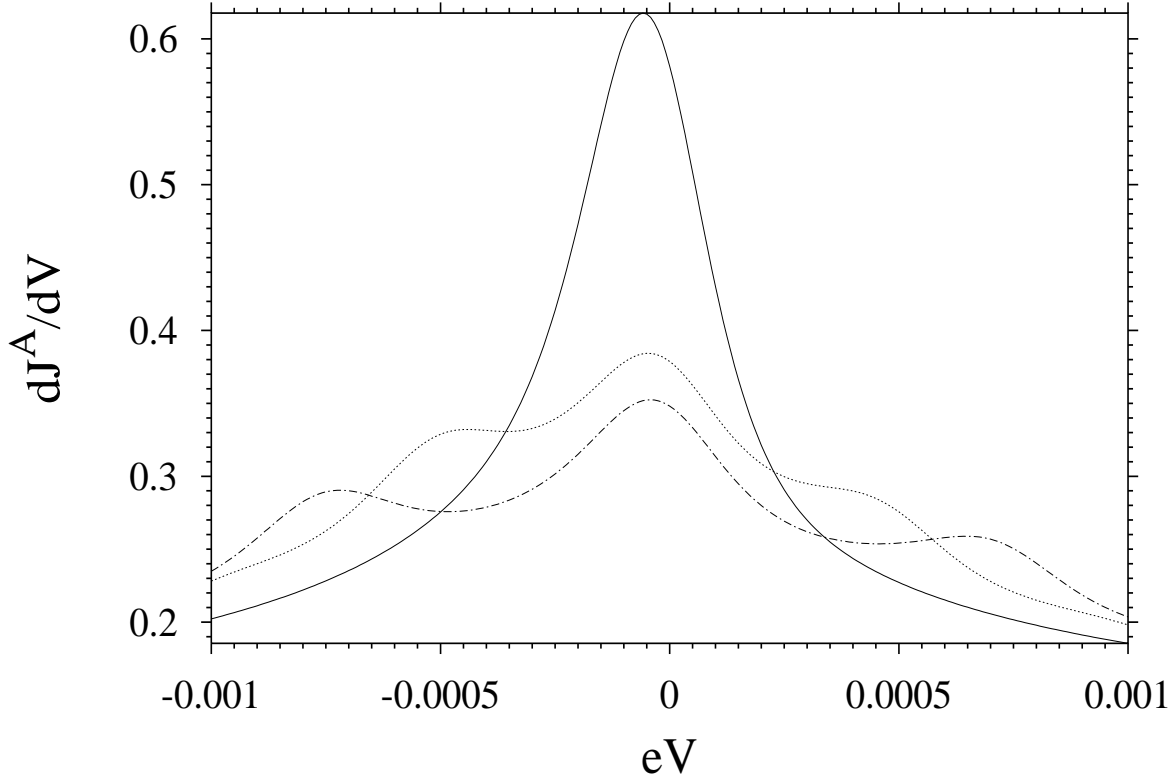}
\caption{
The differential conductance for the {\em N-QD-S} structure
as a function of dc voltage in the units of $4e^2/h$ is plotted 
at the low temperature ($k_BT=1\times 10^{-4}$).
In the absence of the ac field  (sold line), the
differential conductance has the zero-bias peak
due to many-body correlation. 
For  the  ac fields of $\lambda_N=1$ with 
 frequency $\Omega=0.001$ (dotted line) and $\Omega=0.0015$ 
(dash-dotted line), the side peaks at $2eV=n\Omega$ are found 
in addition to the reduced zero-bias peaks.
Here ${\mit \Gamma}^N={\mit \Gamma}^S=0.02$,
$\widetilde \varepsilon_\sigma=\widetilde \varepsilon_{-\sigma}=-0.04$ 
on the scale of  bandwidth $W$.
}
\label{fig3}\end{figure}

%
Differential conductance $dJ^A/dV$ as a function of the dc bias
is shown in Fig.\ref{fig3}. 
The zero-bias peak in the figure originates from the Kondo effect 
associated with the Andreev reflection.
The enhancement of the zero-bias resonance
by the superconductor is already studied \cite{Fazio,kang98-2}.
We find that new small  peaks appear
near the zero-bias peak, 
due to the PAT of the many-body resonance.
The side peaks are located where the voltage is
the multiple of $\mathbf\Omega/2$, instead  $\Omega$, differently from
the case for
{\em N-QD-N} \cite{Hettler,Schiller}.
This implies
that the many-body resonance in the superconducting system changes
from the normal Kondo resonance because of the Andreev reflection.
The Kondo resonance in the superconducting system seems to have
an effective charge $2e$ instead of $e$ due
to the proximity effect.
This proximity coupling seems to allow the side peaks at $2eV=n\Omega$ 
with $n$ being an integer.
\section{Conclusion}
We have studied resonant transport through an interacting quantum
dot coupled to superconductors in the presence of external 
microwave field. We derived a general formula of the current 
for this system, adopting the nonequilibrium Green's function technique.
We have seen that the singularity of BCS density of
states plays an important as well as interesting role
in the quasiparticle PAT.
Influence of microwave fields on the  resonant Andreev transport through
a strongly correlated quantum dot has also been studied.
In addition to a zero-bias peak due to many-body resonance, 
side peaks  are found at the bias voltage corresponding to  
the multiples of the {\em half} of the ac oscillation frequency.

\acknowledgements

This work has been supported in part by 
KOSEF, and STEPI, 
POSTECH/BSRI research fund.
K. Kang acknowledges support by the Max-Planck-Society.



\figure

\end{document}